\documentclass[11pt]{article} % use larger type; default would be 10pt

\usepackage[utf8]{inputenc} % set input encoding (not needed with XeLaTeX)

\usepackage{geometry} % to change the page dimensions
\usepackage{graphicx} % support the \includegraphics command and options

\usepackage{booktabs} % for much better looking tables
\usepackage{array} % for better arrays (eg matrices) in maths
\usepackage{paralist} % very flexible & customisable lists (eg. enumerate/itemize, etc.)
\usepackage{verbatim} % adds environment for commenting out blocks of text & for better verbatim
\usepackage{fancyhdr} % This should be set AFTER setting up the page geometry
\usepackage{sectsty}
\allsectionsfont{\sffamily\mdseries\upshape} % (See the fntguide.pdf for font help)
\usepackage[nottoc,notlof,notlot]{tocbibind} % Put the bibliography in the ToC
\usepackage[titles,subfigure]{tocloft} % Alter the style of the Table of Contents

\usepackage{tikz}         %Tikz for flowchart
\usetikzlibrary{shapes,arrows} %Tikz for flowchart
\tikzstyle{decision} = [diamond, draw, fill=blue!20, 
    text width=4.5em, text badly centered, node distance=3cm, inner sep=0pt]
\tikzstyle{block} = [rectangle, draw, fill=blue!20, 
    text width=7em, text centered, rounded corners, minimum height=4em]
\tikzstyle{line} = [draw, -latex']
\tikzstyle{cloud} = [draw, ellipse,fill=red!20, node distance=3cm,
    minimum height=2em]
    
\usepackage{subfigure} %subsfugure
\title{Layered 2D crystals by design:  optimisation of Sb$_2$Te$_3$--GeTe van der Waals superlattices}
\author{Janne Kalikka, Xilin Zhou, Giacomo Nannicini, and Robert E. Simpson\\ \\ \footnotesize{Singapore University of Technology and Design, Singapore}}
\begin{document}
\maketitle

\section*{Abstract}
\textbf{Herein a genetic algorithm for optimising the design of layered 2D heterostructure is proposed. As a proof-of-concept it is applied to Sb$_2$Te$_3$-GeTe phase-change material superlattices, and the resulting lowest energy structure is grown experimentally. The similarity of the computational and experimental structures is verified with the comparison of XRD spectra. The structure is found to be within 0.92 meV/at. from the energetically most favorable known structure for Ge$_2$Sb$_2$Te$_5$.}

\section{Introduction}
The chalcogenide phase-change materials (PCM) are widely used in data storage due to their unique nonlinear crystallization rate as a function of temperature, and the high optical and electrical contrast between the structural states\cite{burr_phase_2010}. These materials are stable in either phase for years at room temperature, but switch on the nanosecond timescale at elevated temperatures (typically 150-300~$^\circ$C). The material properties, such as crystallization temperature, crystallization rate at elevated temperature, and data retention at room temperature are generally tuned by changing the composition of the material. However, changing the PCM composition changes all of the properties so that fine-tuning one property is likely to affect the others. Recently, it has been reported that it is possible to build layered superlattice structures\cite{simpson_interfacial_2011} of phase-change materials instead of simple alloys. This opens new degrees of freedom for optimization as the properties of these structures can be tuned by changing the layer thicknesses or sequence while keeping the composition constant\cite{ohyanagi_gete_2014}.

Another class of layered materials is the van der Waals heterostructures\cite{geim_van_2013}, which comprise multiple layers of 2D crystals bound together by relatively weak van der Waals forces. Such materials include graphene, hexagonal boron nitride, and molybdenum disulfide. Typically these structures are made by exfoliating the 2D crystals layer by layer, and stacking the layers into the heterostructure one by one. This is currently an active field of research since many material properties are expected to differ in atomically thin materials when compared to bulk. Combining 2D crystals with novel properties to superlattices can lead to vdW materials with designed properties. This opens a similar degree of freedom to materials design as superlattices did for PCM's, only with a wider range of ``building blocks''.

Designing these materials will be a challenge as the material property dependence on the layer sequence is likely to be very complex. Genetic algorithms (GA) have been used successfully in various tasks ranging from protein folding\cite{unger_genetic_1993} to crystal lattice optimization\cite{trimarchi_global_2007}. GA's generally work well in situations where the searched energy landscape has many minima that are separated by significant energy barriers and/or complicated reaction paths. This is because the stochastic nature of GA's means they are less likely to converge to the local minima than deterministic methods\cite{weile_genetic_1997}. The random mutations and recombinations are the heart of GA's. A search starts from a random sample of candidate solutions to the problem, their fitness to the solution is evaluated, and the properties of the best candidates are combined and mutated to create the next generation of candidates.

Layered vdW heterostructures with Ge$_2$Sb$_2$Te$_5$ composition have show to perform higher phase-change efficiency than the alloy material of the same composition\cite{simpson_interfacial_2011}. It is possible that they can be optimised even further. The aim of this paper is to show that GA-led design of 2D heterostructures can be used to find practical structures without the need to survey all the possible structures experimentally. The method is generally applicable to many different problems, and as proof-of-concept it is applied to the optimization of phase-change superlattices. The optimization parameter used is the structure energy since it is relatively simple to calculate, and can be used to assess the stability of the structures grown.

The tellurium sticking coefficient decreases at temperatures above 350~$^\circ$C\cite{chu1997proceedings}. This makes it the highest practical deposition temperature for tellurium superlattices. Therefore a structure is estimated to be sufficiently stable for practical purposes if its energy is less than $k_BT=53.7~\rm{meV/at.}$ above the lowest structure energy found. This will yield a map of potential structures which can then be refined to find a set of desired properties such as optical contrast or resistivity change between states. As different vdW-heterostructures have a wide range of potential applications\cite{geim_van_2013}, we envisage this method impacting on a range of fields from photonics to high-T$_c$ superconductors\cite{orenstein_advances_2000}.

\section{Computational methods}
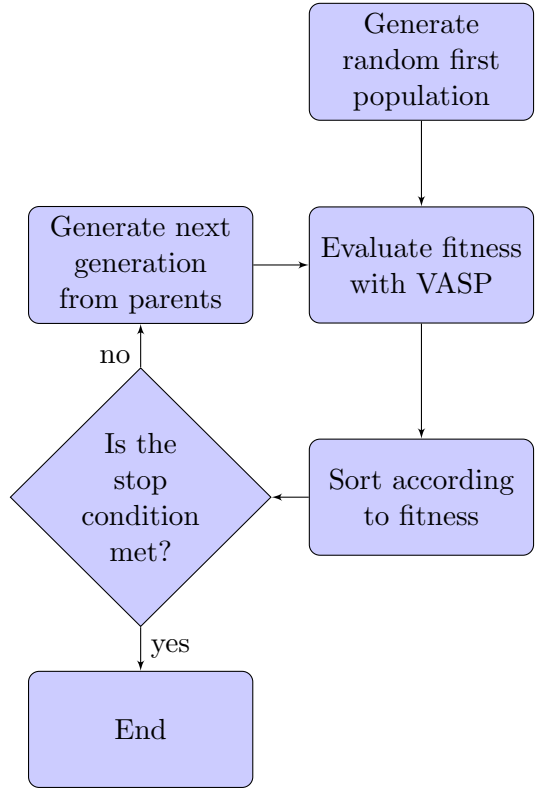
\begin{figure}[htb]
\begin{center}
\begin{tikzpicture}[node distance = 7em, auto]
  \node [block] (init) {Generate random first population};
  \node [block, below of=init] (fitness) {Evaluate fitness with VASP};
  \node [block, below of=fitness,node distance=8em] (sort) {Sort according to fitness};
  \node [block, left of=fitness, node distance=9.6em] (nextgen) {Generate next generation from parents};
  \node [decision, left of=sort,node distance=9.6em] (continue) {Is the stop condition met?};
  \node [block, below of=continue, node distance=8em] (stop) {End};
  
  \path [line] (init) -- (fitness);
  \path [line] (fitness) -- (sort);
  \path [line] (sort) -- (continue);
  \path [line] (continue) -- node [near start] {no} (nextgen);
  \path [line] (continue) -- node {yes}(stop);
  \path [line] (nextgen) |- (fitness);
\end{tikzpicture}
\caption{Flowchart of the algorithm}
\label{flowchart}
\end{center}
\end{figure}

The genetic algorithm procedure is shown in figure~\ref{flowchart}, the first generation of superlattice candidates was randomized, and subsequent generation candidates were combined or mutated from the previous generation. The algorithm uses permutations of 1-$N_{atoms}$ as intrinsic variables. This allows the adaptation of the crossover operator (OX) used in travelling salesman problems, which preserves the relative order of the variables\cite{davis_applying_1985} while combining genes from parent candidates to a child candidate. This speeds up the algorithm since a the order of the atomic layers determines the fitness so preserving good partial sequences has a tendency to produce child candidates with a good fitness. The permutations are mapped to Ge, Sb and Te atomic sequences according to the desired material composition. The program starts by generating a population of permutations, for which it then calculates the fitness function with Vienna Ab initio Simulation Package (VASP)\cite{kresse_ab_1993}. After calculations it reads the fitness values, sorts the population accordingly, and creates the next generation of candidates. The VASP jobs can be executed sequentially, or submitted to a load leveler. In both cases it is possible to run the algorithm automatically up to a specified number of iterations.

The present results were obtained with a population of 20 candidates, each subsequent generation consisted of two elite candidates carried over from the previous generation ``as is'' to preserve the best results throughout the run, 10 candidates produced from previous generation by crossover, and 8 candidates produced with mutation. The possible parent candidates for the next generation were the top 10 candidates.

We used the final energy in the VASP geometry optimization with fixed $X$ and $Y$ coordinates of the atoms, and fixed simulation cell as the fitness function. Essentially the geometry optimization optimized the spacings between the atomic layers while making sure that the order did not change. The simulation cell was hexagonal with $a=b=4.25~\rm{\AA}$, and $c=N_{atoms}\times 1.9372~\rm{\AA}$, which resulted in the experimental crystal phase density of Ge$_2$Sb$_2$Te$_5$. The initial atomic positions were laterally in the fractional coordinates $(0,0)$, $(2/3, 1/3)$, and $(1/3,2/3)$, where the atoms were placed cyclically. The simulations were performed with 240~eV plane-wave cutoff energy, PBEsol\cite{perdew_restoring_2008} exchange-correlation functional, PAW pseudopotentials\cite{kresse_ultrasoft_1999}, periodic boundary conditions, and an 8$\times$8$\times$2 $\Gamma$ centered Monkhorst-Pack \textbf{k}-point grid.

An ordered crossover 1 (OX1) algorithm was used, which is good at preserving the sequence of genes. The parents, p1 and p2, and two cutoff points were selected randomly. The genes of p1 from first cutoff point to the second cutoff point were transfered to the offspring, and the remaining values were taken sequentially from the p2 starting from the gene immediately after cutoff point 2, and omitting the values already present. The mutation algorithm selected a random parent, and swapped two of its genes at random to produce a mutated offspring.

When generating the initial population or the subsequent generations, it was required that the permutation for each candidate resulted in a unique atomic sequence for that generation. For example, if the mutation algorithm would pick an elite candidate, which is carried over to next
generation unchanged, and swap two Te-atoms around in it, the resulting permutation would be rejected since it would result in an identical structure as a permutation already present in that generation. Cyclic permutations and reversed sequences were also rejected, since fitness function calculation used periodic boundary conditions, which deems such structures are identical. Whenever a generated candidate was rejected based on this condition, the generation procedure (randomization, combination or mutation) was repeated until a candidate unique for that generation was generated.

\section{Results}
\begin{figure}[htb]
\begin{center}
\subfigure[Candidate Energies]{\includegraphics[angle=270,width=0.5\textwidth]{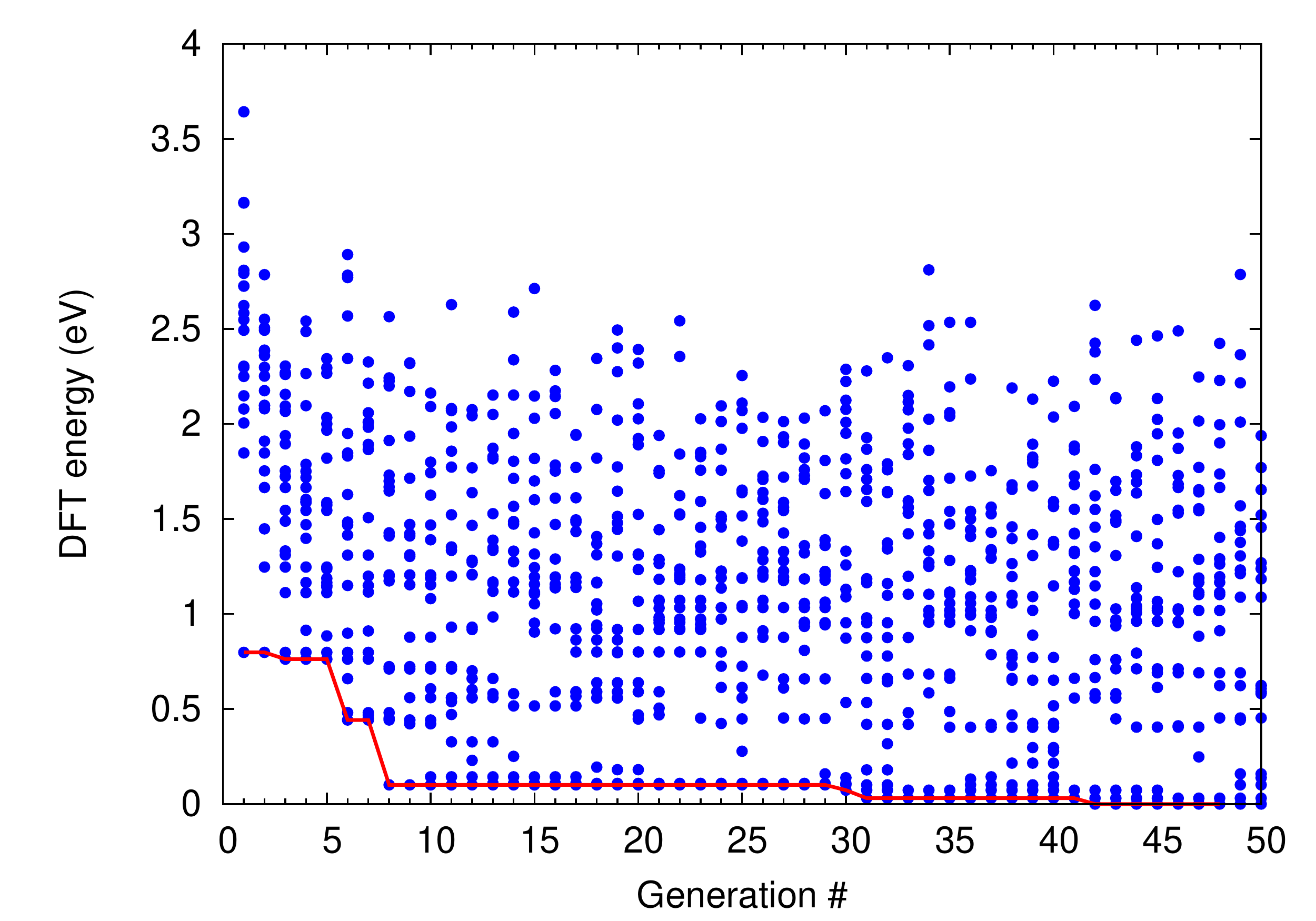}}
\subfigure[Lowest energy candidate structures]{\includegraphics[width=0.5\textwidth]{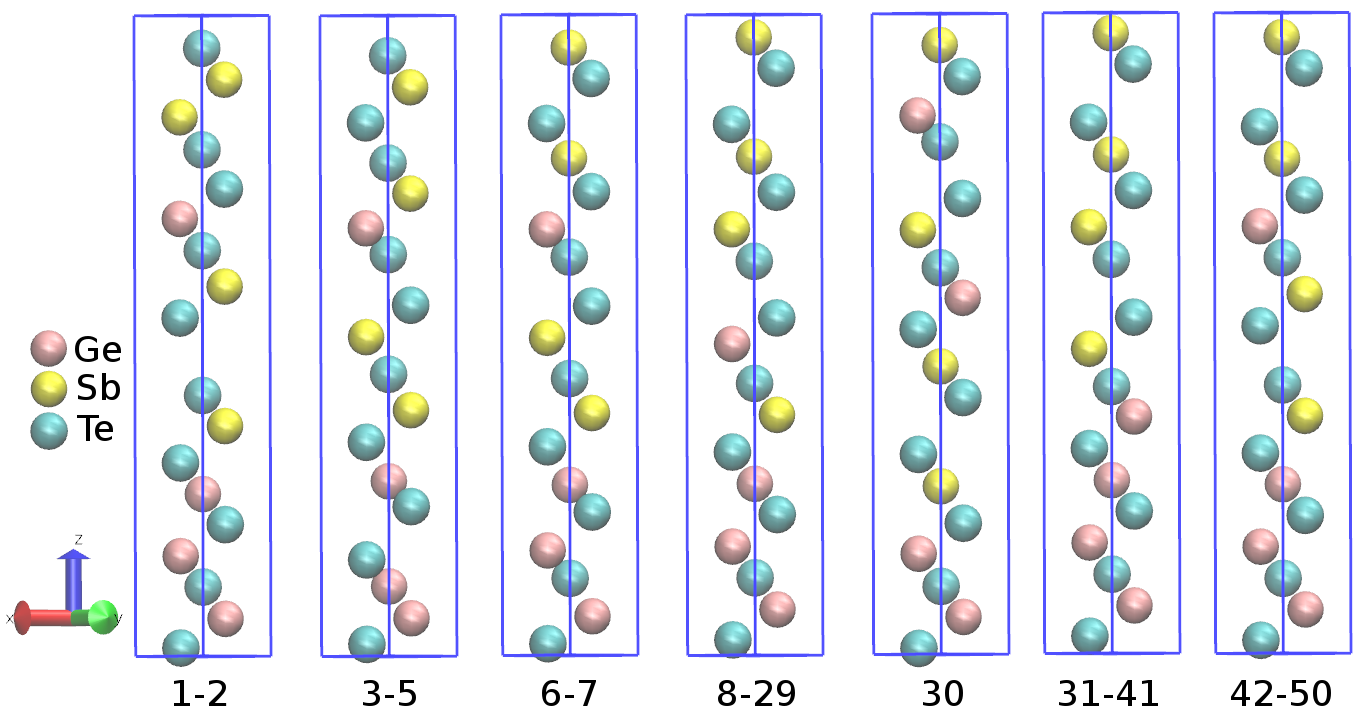}}
\caption{(a) Energies of all candidates as a function of the generation number shown in blue, lowest energy evolution is hilighted in red. (b) Visualizations of the lowest energy structures for each iteration. Colors: germanium in tan, antimony in yellow, tellurium in cyan.}
\label{candidate_energies}
\end{center}
\end{figure}

Figure~\ref{candidate_energies}(a) shows the energies of all candidates for each iteration. The energy of the lowest structure decreases by 0.8~eV over the course of the run, and reaches its lowest value after 42 iterations. The variety of structural energies for each iterations shows that the algorithm is trying structures away from the current minima in order to find better structures. The lowest energy structures are visualized in figure~\ref{candidate_energies}(b), the common features of these are strong A-B-A-B alternation (A: Ge, Sb; B: Te), and separation of GeTe and Sb2Te3. The unavoidable Te-Te vdW interfaces favor Sb neighbours over Ge neighbours.

\begin{figure}[htb]
\begin{center}
\includegraphics[width=0.5\textwidth]{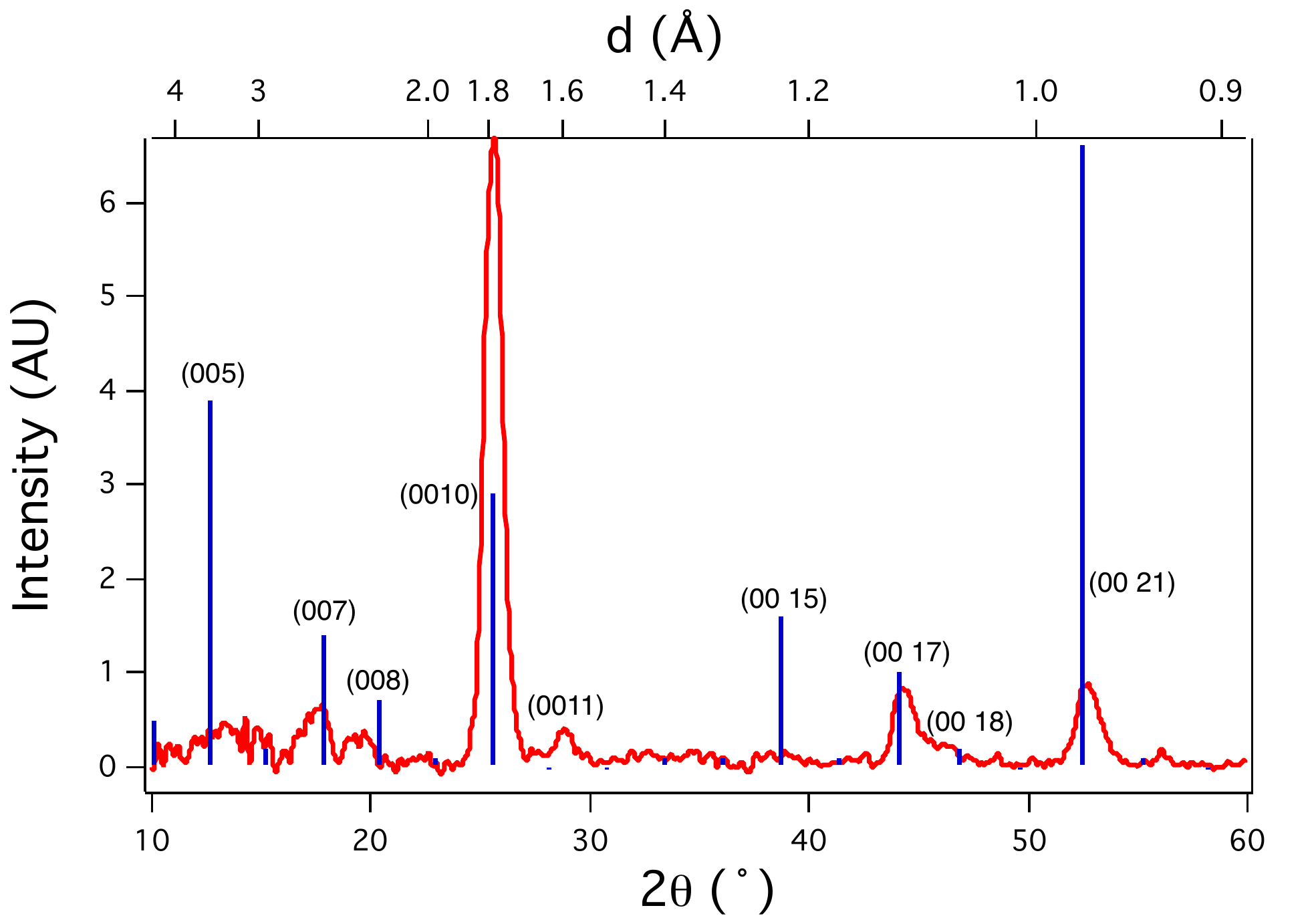}
\caption{Calculated XRD spectra for the lowest energy superlattice structure (blue), and the measured XRD spectra for the sputtered sample (red)}
\label{fig:XRD}
\end{center}
\end{figure}

\begin{figure}[htb]
\begin{center}
\includegraphics[width=0.1\textwidth]{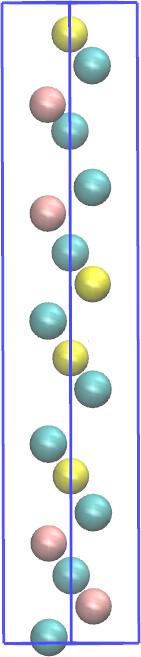}
\caption{The structure of the lowest energy structure with Ge-Te-Te-Ge sequence found. Colors as in fig.~\ref{candidate_energies}(b)}
\label{fig:switchable_structure}
\end{center}
\end{figure}

A structure with the lowest energy is found, and it has an unusual atomic sequence that resembles two layers of quintuple-Sb$_2$Te$_3$ stacked with one and three bilayers of GeTe. This means that the lowest energy structure cannot be modelled with 9 atom unit cell. We were able to grow this sequence onto a Si-substrate, and measure the XRD spectra. Figure~\ref{fig:XRD} shows the comparison between measured XRD spectra, and the spectra calculated from the simulation coordinated using Materials Studio (version 8.0) of Accelrys Inc. The agreement is good, most peaks are present in both spectra, and only one small peak at $\sim$29$^\circ$ is present only in the measurement, and one peak at $\sim$39$^\circ$ is present only in the simulated spectra. In addition to the lowest energy structure we obtained a large number of relatively low energy structures that are therefore practical, and can be investigated further for phase-change, atomic switching, photonics, and memory applications. To elaborate on this, we looked for the lowest energy structure that contains Ge-Te-Te-Ge sequence, as that is regarded as the location where the Ge atomic switching is confined. This structure is referred to as ``switchable'' structure, and it is visualized in figure~\ref{fig:switchable_structure}. This structure is 8.52~meV/at. higher than the lowest energy structure.

\begin{table}
\caption{Energies and simulation cell parameters of different layer orders of Ge$_2$Sb$_2$Te$_5$. The overall lowest energy sequence, and the lowest }\label{tab:structures}
\begin{center}
\begin{tabular}{lrrr}
     &            E (meV/at.) &   a (\AA{})    & c (\AA{})\\
Kooi\cite{kooi_electron_2002} & 0.00 & 4.30 & 34.85\\
Present work & 0.92 & 4.30 & 35.07\\
Ferro GeTe\cite{tominaga_ferroelectric_2014} & 8.12& 4.28 & 35.34\\
Present work (switchable) & 9.44 & 4.28 & 35.36\\
Inverse Petrov\cite{tominaga_ferroelectric_2014} & 13.14 & 4.21& 38.23\\
Petrov\cite{petrov_electron-diffraction_1968} & 17.41 & 4.27&  35.66\\
\end{tabular}
\end{center}
\end{table}

We compared our lowest energy structure with some of the Ge$_2$Sb$_2$Te$_5$ crystal and superlattice structures reported in literature\cite{tominaga_ferroelectric_2014,kooi_electron_2002,petrov_electron-diffraction_1968} by relaxing the respective structures and their simulation cells. Our results are given in table~\ref {tab:structures}, and show that the structures found by GA are very close to the energetically most stable stucture, which is that suggested by Kooi and Hosson\cite{kooi_electron_2002}. The switchable structure is higher in energy by 9.44~meV/at, and is energetically close to the ferro-GeTe structure. It should be noted that the all but inverse Petrov structures have similar lattice constants, and could likely coexist in an imperfect crystal.

\section{Conclusions}

In this paper, a genetic algorithm to efficiently design layer sequences for vdW heterostructures was introduced and applied to Ge$_2$Sb$_2$Te$_5$ superlattices. The algorithm found structures with energies comparable to the energetically most favorable known Ge$_2$Sb$_2$Te$_5$ ground state structures along with ``switchable'' structures which contain Ge-Te-Te-Ge layer sequence. The best structure was grown on a silicon substrate and its XRD pattern agreed well with the spectra computed for the modeled structure. This lends further support that the structures generated by the algorithm are practical.

The fitness function in the algorithm can in practice be any property which can be computed for a structure. This makes it versatile in finding structures or structure combinations that can be used in various applications ranging from photonics to memory devices. The algorithm can be used as a too to help to identify structure--property correlation or structural motifs that can be used to design new van-der-Waals heterostructure superlattices with interesting properties.

\bibliography{refs} 
\bibliographystyle{abbrv} 

\end{document}